\documentclass[letterpaper,twocolumn,10pt]{article}
\usepackage{usenix2019_v3}

\usepackage{graphicx}
\usepackage{amsmath}
\usepackage{amssymb}
\usepackage{booktabs}
\usepackage{array}
\usepackage{float}
\usepackage{enumitem}
\hypersetup{hypertexnames=false}

\makeatletter
\AddToHook{cmd/@makecaption/before}{\nointerlineskip}
\makeatother
\makeatletter
\def\fps@figure{!t}
\makeatother
\begin{document}
\raggedbottom

\date{}

\title{\Large \bf IndexerStage: Staging Indexer-K across Memory Tiers for Sparse-Attention LLM Serving}

\author{%
{\rm Xin Zhang$^{1}$, Weifeng Zhang$^{2}$, Xianyan Xie$^{1}$, Huijun Shen$^{1}$, Xijin Yin$^{1}$, Zhen He$^{1}$}\\
{\rm Yinben Xia$^{2}$, Xiang Li$^{2}$, ZeKun He$^{2}$, Yachen Wang$^{2}$, Xianneng Zou$^{2}$, Guo Chen$^{1,}$\thanks{Corresponding author.}}\\[6pt]
$^{1}$Hunan University \quad $^{2}$Tencent%
}

\maketitle

\begin{abstract}
Long-context sparse-attention serving can remain limited by graphics processing unit (GPU) memory after attention data is offloaded. DeepSeek Sparse Attention retains Indexer-K state to select the tokens that attention will read. Its indexer scans this growing state in full at every layer and decode step. A full scan requires the current layer's state, however, not simultaneous residency of every layer.

We present IndexerStage, a serving runtime that externalizes Indexer-K state while preserving native GPU score, top-$k$, and attention kernels. It stages completed pages through shared execution buffers and preserves new entries locally. Joint memory-and-transfer admission checks storage and each layer's transfer demand before accepting a growing batch. The backing tier gathers pages; the model-serving GPU controls transfer timing and byte budgets.

We implement IndexerStage in SGLang on 16 model-serving and 16 backing H800 GPUs. Across 16K--64K inputs, local streaming storage occupies 0.65--0.70~GiB per model-serving GPU. Relative to host-memory-backed HiSparse, batch size increases by 12.28--23.08\% and decode throughput in tokens per second (TPS) reaches $1.08$--$1.14\times$. At 64K with the latent-KV backing path fixed, the larger batch achieves $1.10\times$ decode TPS; enabling streaming at batch size 24 adds 2.42\% to time per output token. Mixed-length dynamic serving increases mean generating batch size by 12.07\% and delivers $1.31\times$ decode TPS.
\end{abstract}

\begingroup
\setlength{\parskip}{0pt}
\section{Introduction}
\label{sec:intro}

Long-context large language model (LLM) serving can reach a memory limit while larger batches still improve throughput. DeepSeek Sparse Attention (DSA) reduces attention computation by selecting the $k$ highest-scoring token positions, or top-$k$~\cite{liu2025deepseek}. The attention data retained for context tokens form the latent key--value (latent-KV) state. Hierarchical memory keeps most latent-KV state outside the model-serving graphics processing unit (GPU) and fetches the selected entries~\cite{liu2026echo,hisparse,chen2025ess,ma2026sac}. This releases high-bandwidth memory (HBM) on the model-serving GPU.

Selection itself creates another memory limit. DSA's lightning indexer, hereafter the indexer, scores one stored key per visible token before choosing the top-$k$ positions~\cite{liu2025deepseek}. These keys form the \emph{Indexer-K state}. In our 8-bit floating-point (FP8) configuration at 32K input with up to 512 output tokens, this state occupies about 0.25~GiB per request. It accounts for 61.67\% of the request-dependent DSA state remaining after latent-KV offload.

Existing offload systems do not combine externalized active Indexer-K state, native GPU kernels, and commodity hardware. Selective hierarchies offload latent-KV while leaving Indexer-K resident~\cite{liu2026echo,hisparse,chen2025ess,ma2026sac}. DirectKV fetches full-context KV inside a redesigned attention kernel and leaves scheduling policies unchanged~\cite{luo2026no}. KARAT moves the indexer and attention to proposed near-memory hardware~\cite{ham2026heterogeneous}. IndexerStage targets the missing combination for requests that continue decoding.

\emph{A full scan does not require all-layer residency.}

Layers execute in order, so each scan needs its current layer's complete Indexer-K state. During decode, new entries are appended while earlier entries remain unchanged. The earlier entries needed by a scan are known before that layer's query is computed. They can stay in an external \emph{backing tier}, with local HBM reused across layers.

Reusing HBM creates \emph{per-step transfer demand}. The backing tier must supply each offloaded layer's earlier keys again at every decode step. Adding a request therefore requires both storage space and bandwidth to deliver its keys before use.

The difficulty is making released HBM usable by the \emph{running batch}, the requests decoding together. Transfers must arrive before each scan, appends must survive buffer reuse, and movement must limit interference with model execution. We call concurrency supported jointly by storage and a calibrated per-step transfer budget \emph{bandwidth-backed capacity}.

\begin{figure}[!t]
    \centering
    \includegraphics[width=\columnwidth,trim=0 3bp 1bp 0,clip]{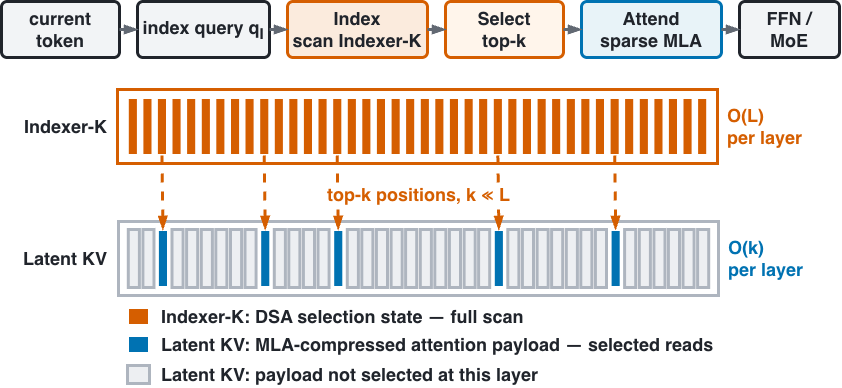}\par
    \caption{The indexer reads $O(L)$ Indexer-K entries to choose the top-$k$ positions. Sparse attention then reads the corresponding $O(k)$ latent-KV entries.}
    \label{fig:sparse-decode-flow}
\end{figure}

We present IndexerStage, the first work to externalize active Indexer-K state with native GPU kernels on commodity hardware. It checks memory and per-step transfer demand before admitting requests, including allowed output growth. Shared buffers hold staged entries; local writable storage preserves appends. Together they form the \emph{local execution view} read by the unchanged indexer.

A \emph{source-side mover} gathers completed entries in the backing tier instead of running the repeated payload gather on the model-serving GPU. The model-serving GPU controls transfer order, timing, and byte budgets, then starts the indexer after all required entries are ready.

Our SGLang prototype uses 16 H800 GPUs for model execution and 16 for backing storage and data movement. Each model-serving GPU connects to a backing device over NVLink/NVSwitch.

The released HBM supports larger batches with a small local streaming allocation. That allocation ranges from 0.65 to 0.70~GiB across the evaluated 16K--64K configurations. At 64K, IndexerStage serves 20.83\% more requests than HiSparse while reducing model-serving HBM use by 1.26\%. Across all three lengths, batch size grows by 12.28--23.08\%, producing $1.08$--$1.14\times$ decode throughput in tokens per second (TPS). A matched latent-KV backing path retains a $1.10\times$ decode-TPS gain at 64K, with a 2.42\% time per output token (TPOT) cost at batch size 24. Under mixed-length arrivals, 12.07\% more generating requests deliver $1.31\times$ decode TPS.

This paper makes four contributions:
\begin{itemize}[leftmargin=*, topsep=0pt, itemsep=0.1em, parsep=0pt, partopsep=0pt]
    \item \textbf{A placement opportunity for full-scan Indexer-K state.} We characterize the HBM limit that remains after latent-KV offload. Each layer scans its complete Indexer-K state, but layer-serial execution lets completed state share temporary HBM rather than remain resident across all layers.
    \item \textbf{Joint memory-and-transfer admission.} We develop a serving runtime that turns cross-layer HBM reuse into bandwidth-backed capacity. It admits a growing running batch only when memory fits and calibrated transfer budgets cover each streamed layer's required state before its scan.
    \item \textbf{A complete local execution view with source-side data movement.} IndexerStage reuses fixed-address execution buffers while preserving new entries and the native score, top-$k$, and attention kernels. The backing tier gathers completed pages; the model-serving GPU controls transfer order, timing, and byte budgets.
    \item \textbf{Implementation and controlled evaluation.} We implement IndexerStage in SGLang on 16 model-serving and 16 backing H800 GPUs. We measure concurrency, TPS, dynamic request handling, and execution costs, with matched-backing-path comparisons separating Indexer-K streaming from the latent-KV backing-path change.
\end{itemize}
\endgroup


\section{Background and Motivation}
\label{sec:background}

\subsection{Inference Background: One Token, Two States}
\label{sec:inference-background}

\subsubsection{Sparse Attention Selects from the Request Context}
\label{sec:background-decode-flow}

Decoding adds one token at a time to each active request. At every layer, Multi-head Latent Attention (MLA) stores the new token's attention data as latent-KV state~\cite{liu2025deepseek}. DSA also stores an Indexer-K entry for selecting which past tokens attention will read.

Selection and attention read different amounts of this growing state. Figure~\ref{fig:sparse-decode-flow} follows one layer's computation: the indexer first scores all $L$ visible Indexer-K entries and returns the top-$k$ positions. Attention then reads the corresponding latent-KV entries. Thus each step reads $O(L)$ Indexer-K entries to choose $O(k)$ latent-KV entries.

\subsubsection{Layer-Serial Execution Still Keeps All Layers Resident}
\label{sec:background-layer-residency}

Layers execute in a fixed order, but their retained state occupies HBM at the same time. Conventional runtimes such as vLLM and SGLang allocate separate state storage for each layer~\cite{vllm-kv-cache-manager,sglang-memory-pool}. When one layer runs, the other layers' state remains allocated. Requests can join, finish, and grow between decode steps without changing that layer order.

The runtime groups a request's Indexer-K entries into fixed-size storage units called \emph{Indexer-K pages}. Each page contains entries for consecutive token positions at one layer. The runtime records where each page is stored; consecutive pages need not occupy adjacent physical addresses.

A page becomes a \emph{completed Indexer-K page} once it is full and receives no further appends. Decode adds entries for new tokens without rewriting earlier entries, so a completed page remains unchanged.

\begin{figure}[!t]
    \centering
    \includegraphics[width=0.92\columnwidth,trim=4bp 10bp 0 8bp,clip]{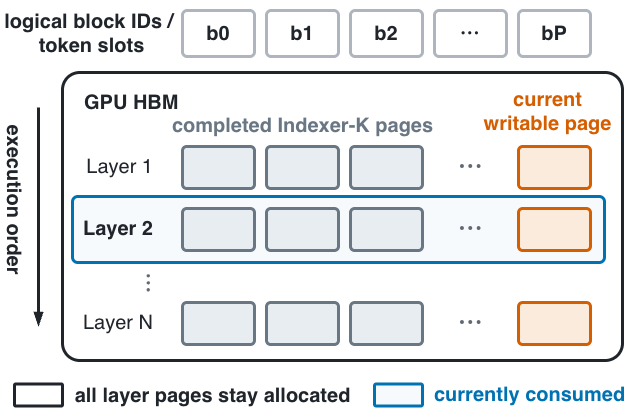}\par
    \caption{Conventional runtimes retain state for every layer while executing one layer at a time. Each request has completed Indexer-K pages and a current writable page at each layer.}
    \label{fig:framework-layer-residency}
\end{figure}

New entries go to the request's \emph{current writable Indexer-K page} at that layer. It stays at a stable address until it fills, after which appends continue on a new page. Figure~\ref{fig:framework-layer-residency} shows how these pages remain allocated across layers even though each layer consumes its state in turn.

\subsubsection{Latent-KV State Sets the First HBM Limit}
\label{sec:background-hbm-anatomy}

Adding requests can improve throughput until their retained state fills HBM. Each decode step then produces more tokens while sharing model weights and execution work.

\begin{figure}[!t]
    \centering
    \includegraphics[width=\columnwidth,trim=27bp 0 24bp 3bp,clip]{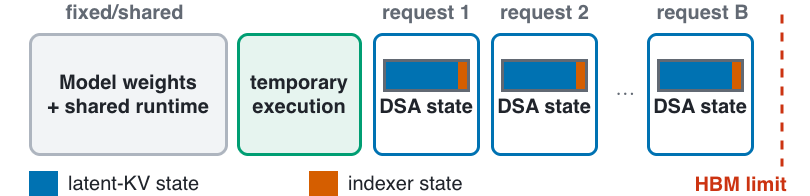}\par
    \caption{HBM allocation for sparse-attention serving. Shared model and runtime storage leaves space for request-dependent latent-KV and Indexer-K state.}
    \label{fig:hbm-growth-model}
\end{figure}

Latent-KV occupies most of the request-dependent DSA state before offload. At 32K input with up to 512 output tokens, latent-KV and Indexer-K occupy 1.24 and 0.25~GiB per request, respectively. Latent-KV accounts for 83.25\% of their combined footprint in this FP8 configuration. Figure~\ref{fig:hbm-growth-model} separates these request-dependent allocations from shared model and runtime storage.

\subsection{Latent-KV Offload Exposes Indexer-K State}
\label{sec:full-scan-bottleneck}

\subsubsection{Latent-KV Offload Moves the Batch Limit}
\label{sec:latent-kv-frontier}

Selective latent-KV access lets hierarchical caches retain most state outside HBM and fetch selected entries~\cite{liu2026echo,hisparse,chen2025ess,ma2026sac}. We use HiSparse, SGLang's hierarchical cache for sparse attention~\cite{hisparse}. It keeps latent-KV state in host dynamic random-access memory (DRAM) and serves attention through a local \emph{latent-KV hot buffer}.

Latent-KV offload raises the batch size and TPS, but the remaining request state still limits concurrency. With 32K input and up to 512 output tokens, it increases batch size from 9 to 39 requests per rank. TPS reaches $2.08\times$ the value with both DSA states resident in GPU HBM. The remaining request state is therefore the next capacity target.

\subsubsection{Indexer-K State Remains Large and Must Be Read in Full}
\label{sec:indexer-hbm-floor}

Indexer-K becomes the larger request-dependent DSA state after latent-KV offload. At 32K and batch size 39, HiSparse retains 0.16~GiB of latent-KV and 0.25~GiB of Indexer-K per request. Indexer-K accounts for 61.67\% of these two resident states. Each layer reads this remaining Indexer-K state in full at every decode step.

\subsubsection{Why Indexer-K State Needs a Different Offload Path}
\label{sec:full-scan-selective-fetching}

Indexer-K offload must support full-context reads at every decode step. Selective latent-KV systems instead move the entries used by sparse attention~\cite{liu2026echo,hisparse,chen2025ess,ma2026sac}. They reduce attention-data movement while leaving the Indexer-K scan in place.

DirectKV fetches full-context KV data inside a redesigned GPU attention kernel~\cite{luo2026no}. Its producer and consumer warps overlap fetching with computation. Applying this fused-fetch path to Indexer-K would require an indexer kernel that consumes remote data. IndexerStage instead prepares the complete local state before the native indexer starts.

KARAT moves computation to the stored state~\cite{ham2026heterogeneous}. Its proposed general-purpose processing-near-memory (PNM) accelerator holds KV and Indexer-K state and executes the indexer and attention. Full Indexer-K reads stay at the storage node rather than crossing the link to a GPU.

Table~\ref{tab:active-state-design-space} compares three properties of the published paths: Indexer-K placement, retention or replacement of native GPU kernels, and use of commodity hardware. The kernel column covers score, top-$k$, and attention. Selective offload refers to the cited latent-KV hierarchies. Their kernel choices differ: ECHO fuses prefetching into the indexer, while HiSparse retains the native indexer~\cite{liu2026echo,hisparse}. IndexerStage externalizes completed pages of streamed layers, while writable state remains local. Each scan still reads the full visible Indexer-K context.

\begin{table*}[!t]
\centering
\setlength{\belowcaptionskip}{4pt}
\caption{Indexer-K placement, GPU-kernel reuse, and hardware requirements of the cited paths. Check marks and crosses describe the published mechanisms. IndexerStage stages completed pages of streamed layers and supplies writable state locally before each full scan.}
\label{tab:active-state-design-space}
\small
\setlength{\tabcolsep}{4pt}
\renewcommand{\arraystretch}{1.15}
\begin{tabular}{@{}>{\raggedright\arraybackslash}p{0.205\textwidth}>{\raggedright\arraybackslash}p{0.335\textwidth}>{\raggedright\arraybackslash}p{0.265\textwidth}>{\raggedright\arraybackslash}p{0.135\textwidth}@{}}
\toprule
\textbf{Mechanism} & \textbf{Active Indexer-K externalized} & \textbf{Native GPU kernels} & \shortstack[l]{\textbf{Commodity}\\\textbf{hardware}} \\
\midrule
Selective offload~\cite{liu2026echo,hisparse,chen2025ess,ma2026sac} & $\times$ --- selected latent-KV entries move; Indexer-K stays local & Varies; ECHO fuses indexer prefetch & $\checkmark$ \\
DirectKV~\cite{luo2026no} & $\times$ --- full-context KV, not Indexer-K & $\times$ --- attention kernel redesigned & $\checkmark$ \\
KARAT~\cite{ham2026heterogeneous} & $\checkmark$ --- Indexer-K and KV at the PNM node & $\times$ --- indexer and attention run on PNM & $\times$ --- proposed PNM \\
\textbf{IndexerStage} & $\checkmark$ --- completed pages staged before each scan & $\checkmark$ --- unchanged & $\checkmark$ \\
\bottomrule
\end{tabular}
\end{table*}

IndexerStage targets full Indexer-K scans with unchanged GPU kernels. The following analysis explains how different layers can share HBM and how the resulting transfers limit the running batch.

\subsection{From Full Scans to Bandwidth-Backed Capacity}
\label{sec:bandwidth-rich-opportunity}

\subsubsection{Layer-Serial Execution Enables HBM Reuse}
\label{sec:layer-residency-opportunity}

Layer-serial execution lets different layers reuse the same HBM. Once a layer finishes reading its completed pages, that temporary space can hold another layer's pages. Figure~\ref{fig:framework-layer-residency} shows the opportunity: conventional allocation retains state even while other layers execute.

A \emph{streamed layer} keeps its completed pages in the backing tier and stages them into reusable HBM before its scan. Their contents no longer change, while the current writable page retains a local location for appends. Other layers remain resident. Temporary capacity depends on the admitted batch and allowed growth, rather than retaining a completed-state allocation for every streamed layer.

\subsubsection{Storage Capacity versus Per-Step Transfer Bandwidth}
\label{sec:hold-vs-feed}

Reusing HBM creates per-step transfer demand for Indexer-K state. Every streamed layer must receive its completed pages again in each decode step. More running requests therefore require both more backing space and more bytes to be transferred before those layers run.

\begin{figure}[!t]
    \centering
    \includegraphics[width=0.76\columnwidth]{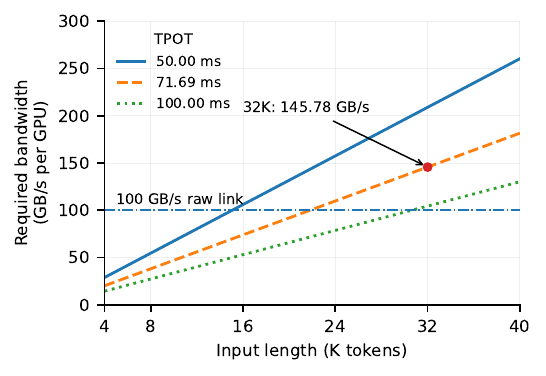}\par
    \caption{Full-state bandwidth estimate at batch size 39 across all 61 layers, including up to 512 output tokens. Each curve uses its indicated TPOT as the entire transfer window. The 32K point uses the measured 71.69~ms HiSparse TPOT. The raw-link line shows nominal bandwidth; overhead and competing traffic reduce the usable share.}
    \label{fig:hold-feed}
\end{figure}

A full-state estimate shows the scale of this demand. Consider batch size 39 per GPU across all 61 layers, with 32K input and up to 512 output tokens. At 520 pages per request per layer and 8,448 bytes per page, the total is 10.45~GB per step. Using the measured 71.69~ms HiSparse TPOT as the entire transfer window yields an idealized average of 145.78~GB/s per GPU. Figure~\ref{fig:hold-feed} shows how this estimate grows with context length.

High-bandwidth paths make this tradeoff practical to explore~\cite{connectx8,cxl4,nvidia-grace-hopper,zuo2025serving}. Raw link bandwidth alone does not determine how many requests can run together. Transfer overhead and competition with model communication can reduce the bandwidth available to Indexer-K. Each layer also needs its state before its scan, so transfer time later in the step cannot cover an earlier shortage.

\subsubsection{Three System Challenges}
\label{sec:streaming-challenges}

Backing capacity supports more running requests only when every layer can read its growing Indexer-K state before use. Three system challenges follow.

\paragraph{C1: Make released HBM usable under the transfer constraint.}
\label{sec:challenge-capacity}
A request may fit in the backing tier while its per-step transfer demand exceeds the bytes available before its layers run. Context growth adds pages even without a new request joining. Admission must therefore check both memory and per-step transfer demand for the entire proposed running batch and its allowed growth.

Transfer timing matters as well as total bytes. A layer can use a destination only after its previous contents are no longer needed. Its required pages must arrive before the scan, so transfer time later in the step cannot cover an earlier shortage.

\paragraph{C2: Reuse HBM while preserving a complete local execution view.}
\label{sec:challenge-view}
Reusing HBM changes where a layer's completed pages are available between scans. Meanwhile, the current writable page continues to receive new entries. The indexer must observe the complete and up-to-date Indexer-K state for each request and layer.

Page transitions must preserve a readable copy of every required entry. A completed page moving to the backing tier remains readable until its replacement is ready and earlier readers finish. Request joins and departures impose the same requirement as mappings change.

\paragraph{C3: Limit interference from moving changing page sets.}
\label{sec:challenge-movement}
The pages needed by a layer change as requests join, finish, and grow. Data movement must track those changes while limiting competition for the model-serving GPU's execution resources and communication bandwidth.

Gathering pages on the model-serving GPU consumes resources also needed by model kernels. A gather kernel can locate and transfer the required pages, but its work grows with the number of page ranges and bytes. The kernel occupies streaming multiprocessors (SMs), the GPU execution units that also run the model kernels.

Host-driven copies avoid gathering payload on those SMs, but require the host to prepare and submit the changing page ranges. The challenge is to support these changes while leaving transfer order, timing, and volume under the model-serving GPU's control.

Completed pages are immutable, writable state has a local owner, and layer order determines when scans need their pages. Page counts make transfer demand known before movement. These properties enable IndexerStage's admission, local execution view, and source-side data movement.

\section{Design}
\label{sec:design}

IndexerStage separates persistent Indexer-K storage from the HBM needed during execution. Layer order enables shared buffers, page counts enable joint memory-and-transfer admission, and immutable completed pages enable source-side data movement. Local writable storage preserves appends throughout these changes.

\subsection{Overview}
\label{sec:design-overview}

IndexerStage chooses placement by layer so all requests follow one staging schedule. Streamed layers keep completed pages in the backing tier; resident layers keep them in model-serving HBM. More streamed layers release more HBM but add transfers at every decode step. The layer set selects this tradeoff, while request lengths determine its transfer demand.

\begin{figure}[!t]
    \centering
    \includegraphics[width=\columnwidth]{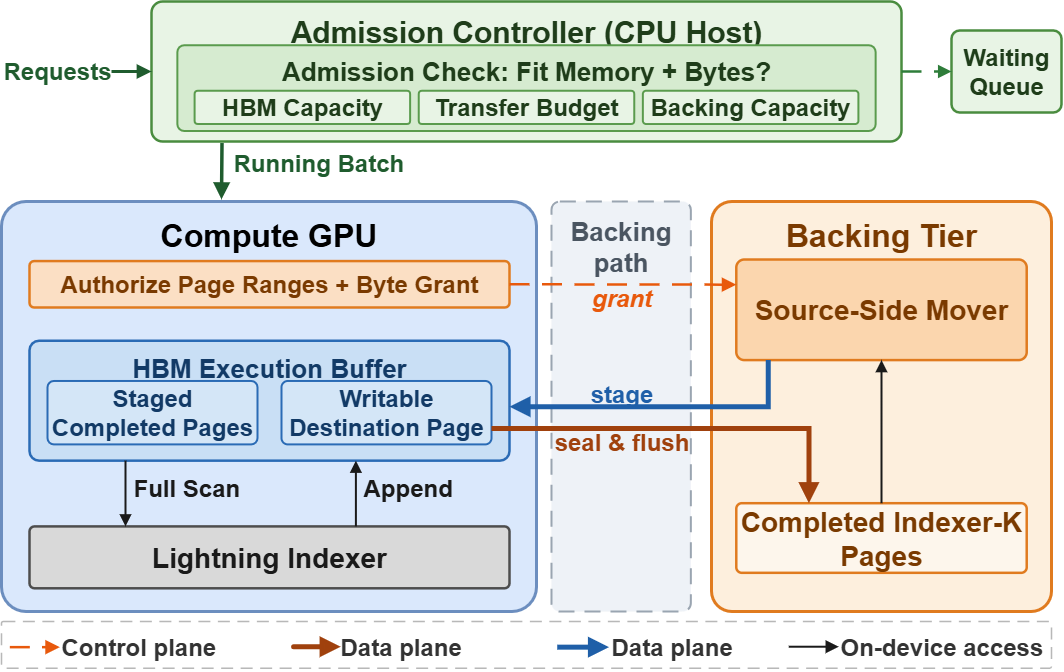}\par
    \caption{Request admission and the Indexer-K data path. Joint memory-and-transfer admission checks storage and per-step transfer demand. The source-side mover stages completed pages into execution buffers, while local writable storage preserves new entries.}
    \label{fig:design-overview}
\end{figure}

Figure~\ref{fig:design-overview} connects request admission to the indexer's local execution view. Joint memory-and-transfer admission checks storage and per-step transfer demand before a request joins the running batch.

Two \emph{fixed-address HBM execution buffers} hold pages staged for streamed layers. Local writable storage preserves new entries between layer uses. The staged pages and writable contents form the local execution view. Source-side data movement fills the buffers under the model-serving GPU's control.
Together, these choices make backing space usable by a growing running batch while preserving the native indexer.

\subsection{Joint Memory-and-Transfer Admission}
\label{sec:capacity-service-coordination}
\label{sec:bandwidth-backed-admission}

IndexerStage checks the whole proposed running batch because requests share storage and transfer opportunities. The controller includes running requests, candidates already selected in the current admission round, and the new candidate. For example, two candidates can each fit the remaining transfer budget when checked alone, yet exceed it when both join. Checking them together prevents this over-admission.

\paragraph{Use page counts to account for transfer demand.}
\label{sec:design-scan-demand}
Page counts expose transfer demand before payload movement. For each streamed layer, the runtime sums the completed pages supplied from the backing tier across the proposed running batch and multiplies by the bytes per page. Summing these bytes across streamed layers gives the per-step transfer demand. Resident pages, locally supplied pages, and unused reservations produce no staging traffic. Admission additionally checks page-rounded output growth.

For example, with 64 entries per page, 130 entries occupy two completed pages and one writable page containing two entries. Once the completed pages have acknowledged backing copies and adopted mappings, staging supplies those two pages; local storage supplies the writable page. Reserved future pages consume capacity but transfer no payload.

\paragraph{Check when bytes can arrive, not just their total.}
\label{sec:design-transfer-budget}
IndexerStage checks transfers in layer order because later bandwidth cannot serve an earlier scan. Calibration assigns a byte budget to each part of model execution that permits new transfers. A layer can use budgets after its destination buffer is released and before its scan starts. Each budget is counted once.

\label{sec:design-joint-admission}
The summed budgets reject batches whose total demand is too large. Ordered checks then ensure each layer's bytes fit before use, accounting for earlier authorized work that can delay transfers. Together with storage checks, this aims to overlap staging with model execution rather than add transfer waits to the decode critical path.

\paragraph{Reserve growth before adding a request.}
\label{sec:design-state-growth}
Admission reserves each request's input plus allowed output length, rounded up to pages. The proposed batch joins execution only if this growth fits storage and transfer budgets. Admission and runtime authorization use the same budgets; completion checks ensure each scan reads every required page even when delivery takes longer than calibrated.

\subsection{A Complete Local Execution View}
\label{sec:state-virtualization}

Different lifetimes call for different storage. Completed pages persist in the backing tier, local writable storage preserves appends, and execution buffers hold temporary copies for the current and next streamed layers. This releases completed-state HBM without losing newly generated entries.

\paragraph{Reuse two buffers across layers.}
\label{sec:design-bounded-view}
Two execution buffers allow staging to overlap an indexer scan without overwriting the pages being read. Buffer~0 can serve one streamed layer while Buffer~1 prepares the next. One buffer would serialize its refill with its current scan. A separate full buffer per streamed layer would retain all-layer residency. Two buffers retain overlap while sharing completed-page space across layers.

A buffer becomes reusable after its scan finishes and appended contents have been preserved. Resident layers use their own allocations, so the next streamed layer need not be adjacent. Requests share buffer capacity according to context length and reserved growth.

\begin{figure}[!t]
    \centering
    \includegraphics[width=\columnwidth]{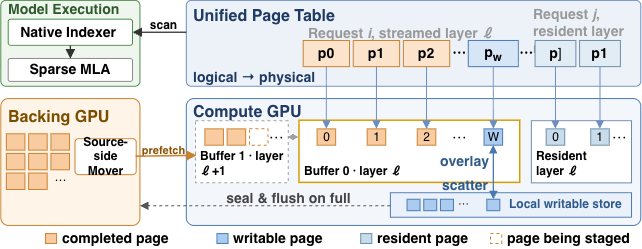}\par
    \caption{The local execution view for streamed layer $\ell$ combines staged completed pages with writable-page contents. It also retains any completed local page whose backing copy is not yet adopted. Buffer~1 prepares the next streamed layer.}
    \label{fig:execution-view}
\end{figure}

\paragraph{Preserve the native page-table interface.}
\label{sec:design-page-table}
IndexerStage changes placement rather than the native page-table interface. Figure~\ref{fig:execution-view} combines staged completed pages and writable contents in logical token order. Appends and scans refer to the same writable destination; the other buffer prepares a later layer.

The unchanged indexer reads the same entries and returns top-$k$ token positions. Attention resolves those positions through its existing latent-KV mapping. Exactness means preserving the native indexer's input and selection semantics.

\paragraph{Keep a valid copy throughout page handoff.}
\label{sec:design-state-separation}
Page handoff retains a valid copy until its replacement is ready and earlier reads finish. When a page fills, the runtime \emph{flushes} it to the backing tier and sends subsequent appends to another writable destination. The local copy remains readable until the backing copy and its mapping can be adopted together.

Request changes use the same rule. A joining request receives a complete local execution view before activation. A departing request disappears from future views, while earlier accesses retain their storage until safe reuse.

\subsection{Source-Side Data Movement}
\label{sec:source-executed-dataplane}

\begin{figure}[!t]
    \centering
    \includegraphics[width=\columnwidth]{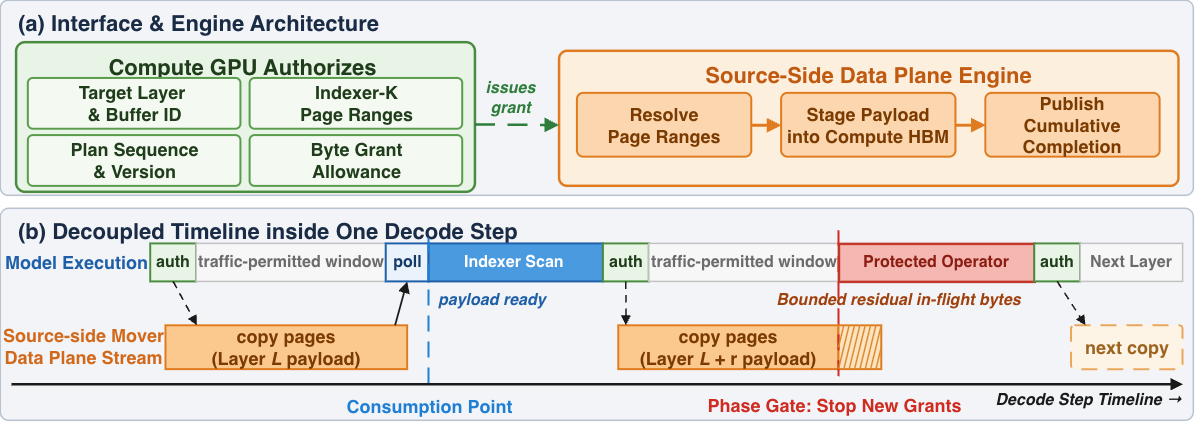}\par
    \caption{Source-side data movement. Panel~a separates model-serving-GPU authorization from source-side payload gathering. Panel~b places new authorization at transfer-permitted boundaries and checks completion before the indexer reads the destination.}
    \label{fig:authorization}
\end{figure}

Gathering runs at the backing tier to leave model-serving GPU execution resources to the model. The model-serving GPU controls transfer order, timing, and byte budgets because it knows when buffers are free and communication-sensitive operators run. Figure~\ref{fig:authorization} shows this separation of payload gathering and control.

\paragraph{Separate changing pages from the layer schedule.}
\label{sec:design-logical-plans}
Request changes update the needed pages without changing the model's layer order. IndexerStage prepares those mappings before execution, leaving the per-layer control path to authorize prepared work. Each authorization names a target layer and execution buffer and limits the bytes that may be moved. The source-side mover gathers the corresponding completed pages. This avoids submitting one host copy for every page at every layer.

\paragraph{Transfer the current payload.}
\label{sec:design-movement-separation}
The mover follows current page mappings even when backing storage is fragmented. Unused reservations create no payload, and pages awaiting flush completion remain locally supplied. Different request lengths therefore use the same schedule without transferring padding.

\paragraph{Use phase gates together with byte grants.}
\label{sec:design-phase-budget}
Source-side gathering still consumes interconnect bandwidth and can delay communication-sensitive model operators. Phase gates choose when new transfers may start; byte grants limit how much work they authorize. Gates alone can issue excessive bytes, while byte limits alone can select disruptive times. Admission also accounts for bounded work already authorized before a phase boundary.

Authorization starts movement; completion permits consumption. The indexer waits until every transfer for its current local execution view is complete.

\section{Implementation}
\label{sec:implementation}

The SGLang prototype implements source-side data movement with a C++/CUDA extension over NVLink/NVSwitch. Startup allocates storage and builds the calibrated transfer schedule. Request preparation supplies mappings to the captured decode graphs.

\subsection{Paged Storage and Native Interfaces}
\label{sec:implementation-state}

IndexerStage preserves the native Indexer-K page layout. Each entry contains a 128-byte FP8 vector and a 4-byte scale. A page stores 64 vectors followed by their scales, totaling 8,448 bytes. The source-side mover copies this layout directly into an execution buffer.

CUDA Graph capture requires fixed-address HBM execution buffers and metadata. Request changes update mappings without reallocating graph-visible storage. Both execution buffers have equal size and the same four regions: shared completed-page space, per-request context slack, decode-growth reservations, and writable destinations. Their page counts follow the configured cache capacity and request slots. Separate local writable storage preserves one page per slot and streamed layer between uses.

The prototype allocates 256-page context segments and, by default, 32-page growth chunks from shared pools. Source and destination chunks have independent physical base addresses. Per-request metadata records chunk locations and completed-page counts. Growth reservations use capacity within these pools; unassigned space still occupies physical memory.

SGLang's Indexer-K page table directs native appends and scans to the same writable destination. Latent-KV retains its separate request-to-token mapping, so top-$k$ positions still select the original attention data.

\subsection{Transfer-Budget Calibration}
\label{sec:implementation-control}

Calibration measures the parts of model execution available for Indexer-K transfers. It profiles a reference configuration with resident Indexer-K state and the same peer-HBM latent-KV path. Configured latent-KV loading, expert dispatch, and combine phases are excluded from new transfer authorization.

The builder multiplies median permitted-phase durations by the lowest byte-weighted median staging bandwidth across ranks. Using the slowest measured rank avoids budgeting for faster ranks and delaying them at synchronization. A configurable margin scales the bandwidth, and budgets are rounded down to whole pages.

The table records the model, page format, streamed layers, device configuration, protected phases, and maximum context. Mixed-length batches use that context's table with their current page counts and growth reservations. Runtime waiting never increases the calibrated budgets.

\subsection{Request Admission and Batch Preparation}
\label{sec:implementation-integration}

The scheduler applies Section~\ref{sec:capacity-service-coordination} before activating requests. It checks request slots, the latent-KV hot buffer, local and backing pools, metadata capacity, and supported CUDA Graph batch sizes. Prompt length plus allowed output growth is capped at the model limit. A candidate stays queued if memory or transfer checks fail.

A background central processing unit (CPU) worker builds affected page tables and transfer descriptors while decode continues with the current plan. It reuses unchanged mappings. The model-serving GPU gathers missing page identifiers from SGLang's request-to-token table. Joining requests receive backing copies and initialized local writable storage.

At a replay boundary, the runtime adopts the prepared page tables, append destinations, descriptors, and transfer plan as one version after state preparation and metadata uploads finish.

\subsection{Decode Execution and Request Updates}
\label{sec:implementation-graph}
\label{sec:implementation-mover}

Each decode step launches a task graph on the backing device over prepared page ranges. Graph hooks on the model-serving GPU authorize transfers at calibrated execution points. The gather kernel copies the authorized bytes and publishes completion with system-scope acquire/release ordering. Version checks bind completion to the current plan.

The acquire hook waits for the layer's final staging command, then overlays saved writable contents into its execution buffer. Native store, score, and top-$k$ kernels use the resulting page table.

The release hook runs a scatter kernel on the model-serving GPU to save new entries in local writable storage. When a page fills, it flushes that page to the backing device and switches appends to the alternate destination. The buffer-release event follows preservation. After flush acknowledgment, a later replay adopts the backing source and destination mapping together.

Request retirement removes a request from subsequent replays and reclaims its storage after model operations and backing tasks complete. An inactive execution slot can remain available for reuse without committing a batch-removal plan, separating frequent local cleanup from shared-plan changes.



\section{Evaluation}
\label{sec:evaluation}

We evaluate whether Indexer-K offload increases useful serving capacity after accounting for its storage and execution costs. Static experiments measure the additional requests and TPS, then explain the resulting memory use. Dynamic experiments test whether the gains persist as requests join, grow, and finish, and measure the work required to update their state. Varying layer placement exposes the tradeoff between released HBM and per-step transfers. A matched latent-KV backing path then separates the benefit of Indexer-K offload from faster attention-data access.

\subsection{Experimental Setup}
\label{sec:eval-setup}

\begingroup
\makeatletter
\renewcommand\paragraph{\@startsection{paragraph}{4}{\z@}%
    {5pt plus 1pt minus 1pt}{-1em}{\normalfont\normalsize\bfseries}}
\makeatother

\paragraph{Testbed and runtime.}
We use four servers with eight NVIDIA H800 GPUs each. Four GPUs per server execute the model, and four form the backing tier. Each model-serving GPU connects to one device in the backing tier over NVLink/NVSwitch and provides 79.10~GiB of usable HBM. We run the 61-layer DeepSeek-V3.2 model with FP8 state and 16 data-parallel (DP) attention ranks. The score, top-$k$, and attention kernels are unchanged. The dynamic runs use a shared SGLang-based runtime derived from commit~\texttt{1519acf}.

\paragraph{Baselines.}
Our \textsc{HiSparse} baseline keeps Indexer-K state in HBM on the model-serving GPU and stores latent-KV state in pinned host DRAM. \textsc{IndexerStage} uses peer HBM for latent-KV state and completed Indexer-K pages from selected layers. We also implement \textsc{LatentPool-HBM}, which shares IndexerStage's latent-KV access and writeback path but keeps Indexer-K state resident. It isolates Indexer-K streaming in Section~\ref{sec:eval-backing-path}. The dynamic configurations use the same SGLang-derived codebase with their respective placement, writeback-slot, and admission settings.

\textsc{Memory-Only} retains IndexerStage's placement and mover but disables transfer-demand checks during admission. It proportionally increases runtime byte budgets while keeping command plans, issue points, and completion checks unchanged.

\paragraph{Relationship to Table~\ref{tab:active-state-design-space}.}
HiSparse represents selective latent-KV offload~\cite{hisparse}. DirectKV targets attention rather than the DSA indexer and leaves scheduling policies unchanged; it provides no bandwidth-based admission rule~\cite{luo2026no}. KARAT instead relies on its proposed PNM accelerator~\cite{ham2026heterogeneous}. Externalizing a full Indexer-K scan still requires its bytes to cross the backing path. Memory-Only isolates the missing admission check within IndexerStage's otherwise unchanged staging path.

\paragraph{Backing-path bandwidth.}
Peer HBM provides headroom for full-context Indexer-K transfers. The full-step estimate in Figure~\ref{fig:hold-feed} is 145.78~GB/s at 32K and batch size 39, or 47.80~GB/s for the 20 layers streamed at $r=3$. Against a 50.00~GB/s usable host-path assumption, that leaves 2.20~GB/s before selective latent-KV traffic and before-scan timing constraints. At 64K, batch size 24 and a 62.66~ms TPOT give 66.79~GB/s. These demands motivate the high-bandwidth tier; Section~\ref{sec:eval-backing-path} measures the latent-KV path-only gain separately.

\paragraph{Layer placement.}
The offload stride $r$ selects every $r$th layer for Indexer-K streaming. Unless stated otherwise, $r=3$: layers 3, 6, $\ldots$, 60 are streamed, and 41 layers remain resident. The 20 streamed layers receive their completed pages at every decode step. Section~\ref{sec:eval-admission} varies $r$ from 1 to 4.

\paragraph{Workloads and metrics.}
Static experiments use real KV state generated from synthetic prompts and imported before timed decode. This emulates the decode side of a prefill--decode-separated deployment. Input lengths are 16K, 32K, and 64K, with up to 512 output tokens. Dynamic experiments mix request lengths and arrival times, as described in Section~\ref{sec:eval-dynamic}.

Batch size is per DP rank. Decode TPS counts generated tokens across model-serving ranks during decode execution; prefill computation is excluded. Static runs report average TPOT, while dynamic runs report percentiles across requests. Memory uses GiB, with $2^{30}$ bytes per GiB; transfer volumes and rates use GB and GB/s.

\par\endgroup

\subsection{Static Throughput and Memory Use}
\label{sec:eval-e2e}
\label{sec:eval-static}

\subsubsection{Capacity and Decode Throughput}
\label{sec:eval-static-throughput}

\begin{figure}[!t]
    \centering
    \includegraphics[width=0.88\columnwidth]{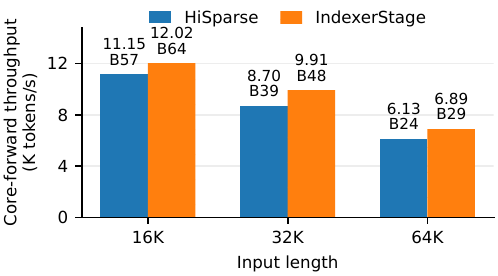}\par
    \caption{Static decode performance at $r=3$. Lines show decode TPS on the left axis; bars show maximum admitted batch size per rank on the right axis.}
    \label{fig:eval-static}
\end{figure}

IndexerStage increases batch size by 12.28--23.08\% and achieves $1.08$--$1.14\times$ the TPS of HiSparse across 16K--64K inputs. Figure~\ref{fig:eval-static} shows that the released memory supports additional requests at every input length. Each system runs at its maximum admitted batch size.

The additional requests outweigh the TPOT increase at both 16K and 32K. At 16K, batch size increases from 57 to 64, or 12.28\%. Decode TPS rises from 11,153.80 to 12,019.60, an increase of 7.76\%, while TPOT rises by 4.18\%, from 81.77 to 85.19~ms. At 32K, batch size increases from 39 to 48, or 23.08\%. Decode TPS rises from 8,704.20 to 9,913.60, an increase of 13.89\%, while TPOT rises by 8.06\%, from 71.69 to 77.47~ms.

The additional capacity also improves decode TPS at 64K. Batch size increases from 24 to 29, or 20.83\%. Decode TPS rises from 6,128.10 to 6,886.70, an increase of 12.38\%, reaching $1.12\times$ the HiSparse result. TPOT rises by 7.53\%, from 62.66 to 67.38~ms. In all three cases, each iteration produces proportionally more tokens than the additional time it takes.

Decode TPS falls for both systems as the input grows because longer contexts leave room for fewer requests and require more Indexer-K reads per request. Offloading completed Indexer-K state gives IndexerStage room for more requests at each length. Its TPS gains are smaller than its batch-size gains because the larger batches require more scoring, sparse attention, expert communication, and staging.

\subsubsection{Memory Released for Additional Requests}
\label{sec:eval-memory}
\label{sec:eval-memory-placement}

At 64K, IndexerStage serves 20.83\% more requests per rank while reducing total HBM use by 1.26\%. Table~\ref{tab:eval-memory} shows total usage falling from 78.10 to 77.11~GiB as batch size rises from 24 to 29. The remaining capacity cannot admit another complete request under the same storage and growth-reservation checks.

\begin{table}[t]
    \centering
    \footnotesize
    \setlength{\tabcolsep}{2pt}
    \begin{tabular*}{\columnwidth}{@{}l@{\extracolsep{\fill}}rrr@{}}
        \toprule
        & HiSparse & IndexerStage & Change \\
        Batch size per rank & 24 & 29 & +20.83\% \\
        \midrule
        \multicolumn{4}{l}{\textit{Per-request resident state (GiB)}} \\
        Resident-layer Indexer-K state & 0.51 & 0.34 & $-32.79\%$ \\
        Latent-KV hot buffer & 0.16 & 0.16 & 0.00\% \\
        Indexer-K share & 76.30\% & 68.40\% & $-10.35\%$ \\
        \midrule
        \multicolumn{4}{l}{\textit{HBM on the model-serving GPU (GiB)}} \\
        Resident-layer Indexer-K state & 12.16 & 9.88 & $-18.79\%$ \\
        Execution buffers & 0.00 & 0.65 & \\
        \shortstack[l]{Writable storage +\\ metadata/workspace} & 0.00 & 0.03 & \\
        Total HBM usage & 78.10 & 77.11 & $-1.26\%$ \\
        \midrule
        \multicolumn{4}{l}{\textit{Backing pools (GiB per paired rank)}} \\
        Medium & DRAM & HBM & --- \\
        Latent-KV payload & 58.62 & 70.83 & +20.83\% \\
        Indexer-K pages + metadata & 0.00 & 4.82 & --- \\
        Latent-KV headroom & 1.83 & 2.44 & +33.04\% \\
        Total listed pool allocation & 60.45 & 78.09 & +29.18\% \\
        \bottomrule
    \end{tabular*}
    \caption{64K memory at batch sizes 24 and 29. Changes are relative to HiSparse and use values before display rounding. Indexer-K share uses resident-layer Indexer-K state plus the latent-KV hot buffer as its denominator. Snapshots use the highest-memory rank in each run. GPU totals include model and runtime allocations; backing totals cover the listed pools. Dashes mark undefined percentage comparisons.}
    \label{tab:eval-memory}
\end{table}

The resident-state reduction comes from offloading completed Indexer-K pages from 20 layers. The $r=3$ placement streams every third layer to limit per-step traffic under the transfer budget. Resident-layer Indexer-K state per request falls by 32.79\%, from 0.51 to 0.34~GiB. Even with the larger batch, this allocation falls by 18.79\%, from 12.16 to 9.88~GiB per GPU. The latent-KV hot buffer stays at about 0.16~GiB per request, so its total footprint grows by 20.83\% with batch size.

Local streaming storage occupies at most 0.70~GiB at all three capacity points: 0.65~GiB at 16K, 0.70~GiB at 32K, and 0.68~GiB at 64K. At 64K, the two equal execution buffers occupy 0.65~GiB; writable storage and metadata/workspace add 0.03~GiB. The buffers contain completed-page space, context slack, growth reservations, and writable destinations. Their sizes follow the configured cache capacity and request slots. Longer contexts coincide with smaller batches in this sweep, keeping the total within a narrow range.

The additional requests also make more use of the backing tier. Their latent-KV payload grows by 20.83\%, from 58.62 to 70.83~GiB, while offloaded Indexer-K pages and metadata occupy another 4.82~GiB. Thus the larger batch retains more state outside the model-serving GPU even as its total local HBM use falls.

\subsection{Online Decode with Dynamic Arrivals}
\label{sec:eval-dynamic}
\label{sec:eval-dynamic-arrivals}

\subsubsection{Mixed-Length Workload}
\label{sec:eval-dynamic-workload}

Dynamic serving exercises IndexerStage as requests join, grow, and finish. Both systems receive the same Poisson arrivals at 12 requests/s, routed uniformly across the 16 DP ranks. Input lengths are 16K, 32K, 48K, and 64K; output lengths are 256, 512, 768, and 1,024 tokens. Each length accounts for approximately one quarter of its respective distribution. IndexerStage uses $r=3$ and the transfer-budget table for the workload's maximum configured context.

We measure 8,192 requests after 1,440 warm-up requests. Measurement requests arrive over 674.26~s and are tracked through completion. Prepared request state replaces timed prefill computation, as in the static experiments.

\subsubsection{Throughput and Token Latency}
\label{sec:eval-dynamic-results}

IndexerStage serves more concurrent requests with lower token latency. Table~\ref{tab:eval-dynamic} shows $1.31\times$ decode TPS as mean generating batch size rises from 24.27 to 27.20. Median TPOT falls by 12.95\%, and both tail percentiles fall by more than 12.00\%.

\begin{table}[t]
    \centering
    \footnotesize
    \setlength{\tabcolsep}{2pt}
    \begin{tabular*}{\columnwidth}{@{}l@{\extracolsep{\fill}}rrr@{}}
        \toprule
        Metric & HiSparse & IndexerStage & Change \\
        \midrule
        TPS & 4,274.97 & 5,611.49 & $1.31\times$ \\
        Mean generating batch size & 24.27 & 27.20 & $1.12\times$ \\
        TPOT p50 (ms) & 91.75 & 79.86 & $-12.95\%$ \\
        TPOT p95 (ms) & 95.11 & 83.39 & $-12.32\%$ \\
        TPOT p99 (ms) & 98.91 & 85.82 & $-13.23\%$ \\
        \bottomrule
    \end{tabular*}
    \caption{Dynamic serving on the same arrival sequence. Mean batch size counts token-producing requests per rank over the full run. TPOT percentiles cover the 8,192 measurement requests. Changes are ratios for TPS and batch size, and relative reductions for TPOT.}
    \label{tab:eval-dynamic}
\end{table}

Writeback overlap also helps the dynamic execution path. In the HiSparse configuration of our SGLang version, each request reuses one slot for newly generated latent-KV entries and waits if its previous writeback is unfinished. IndexerStage alternates between two slots, allowing writeback to overlap use of the other slot. Section~\ref{sec:eval-backing-path} controls for the shared access and writeback path when isolating Indexer-K streaming.

The benefit extends across request lengths and output growth. IndexerStage reduces TPOT for 8,159 of the 8,192 paired requests, or 99.60\%. Median paired reductions are 11.44--11.48~ms across input lengths and 11.18--11.78~ms across output lengths. Both systems complete every measurement request, so the gains persist as batch membership changes.

\subsubsection{Cost of Updating Request State}
\label{sec:eval-dynamic-state-cost}

The foreground commit takes a few milliseconds for most measured state updates. Table~\ref{tab:eval-dynamic-events} reports runtime event timings pooled across all 16 ranks over the full dynamic run. The median commit takes 3.00~ms for a single-request join and 3.19~ms for a page-growth update. Their 95th percentiles are 4.22 and 4.15~ms. Page-growth updates can include several requests, rather than one physical page.

\begin{table}[t]
    \centering
    \footnotesize
    \setlength{\tabcolsep}{3pt}
    \begin{tabular*}{\columnwidth}{@{}l@{\extracolsep{\fill}}rrr@{}}
        \toprule
        Timed operation & Samples & \shortstack{p50\\(ms)} & \shortstack{p95\\(ms)} \\
        \midrule
        Single-request join commit & 3,752 & 3.00 & 4.22 \\
        Request-local retirement & 9,616 & 0.33 & 1.29 \\
        Single-request removal commit & 561 & 3.03 & 4.07 \\
        Page-growth update commit & 3,669 & 3.19 & 4.15 \\
        \bottomrule
    \end{tabular*}
    \caption{IndexerStage state-update timings over the full dynamic run, including warm-up and drain. Join and removal rows time plan commit; page growth times the enclosing update commit. Request-local retirement includes cleanup and logging. Rows describe separate events and are not additive.}
    \label{tab:eval-dynamic-events}
\end{table}

Frequent cleanup avoids a shared-plan commit. As described in Section~\ref{sec:implementation-graph}, finished requests can retire locally while their execution slots await reuse. Median local retirement takes 0.33~ms, versus 3.03~ms for a single-request removal commit.

Plan construction runs on the CPU while decode uses the current plan. Its median durations range from 28.64 to 68.01~ms across removals, joins, and page growth. Preparing this work in the background leaves mapping adoption at the replay boundary, explaining the shorter foreground commits.

\subsection{Admission under Memory and Transfer Limits}
\label{sec:eval-admission}

\begin{figure}[!t]
    \centering
    \includegraphics[width=0.76\columnwidth]{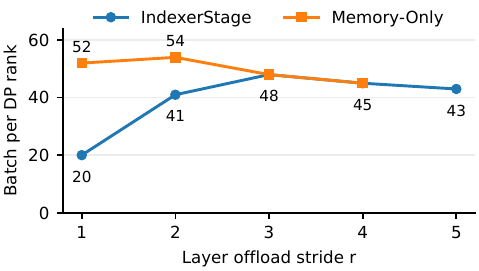}\par
    \caption{Admitted batch size per rank at 32K. Smaller $r$ streams more layers. IndexerStage checks both memory and transfers; Memory-Only admits requests by memory capacity and expands runtime byte budgets.}
    \label{fig:eval-admission}
\end{figure}

Figure~\ref{fig:eval-admission} shows a change from a transfer-limited batch to a memory-limited batch. We vary $r$ from 1 to 4 at 32K and compare IndexerStage with Memory-Only at each layer placement. Increasing $r$ streams fewer layers, reducing each request's per-step transfer demand while retaining more Indexer-K state in local HBM.

Memory-Only varies within a narrower range, from 45 to 54 requests. Changing $r$ shifts persistent Indexer-K state between local HBM and the backing tier rather than eliminating it. Its capacity checks reflect that placement, but omit the cost of supplying the state before every scan. IndexerStage instead rises from batch size 20 at $r=1$ to 41 at $r=2$ and 48 at $r=3$. Fewer streamed layers reduce the repeated bytes per request, allowing the transfer checks to accept more requests. At $r=3$, the two curves meet. Increasing $r$ further to 4 leaves more layers resident and both batch sizes fall to 45. Beyond the meeting point, reduced transfer demand no longer compensates for the extra local residency.

\begin{table}[t]
    \centering
    \footnotesize
    \setlength{\tabcolsep}{3pt}
    \begin{tabular*}{\columnwidth}{@{}c@{\extracolsep{\fill}}rrrr@{}}
        \toprule
        & \multicolumn{2}{c}{TPS (K tokens/s)} & \multicolumn{2}{c}{Memory-Only wait (ms)} \\
        \cmidrule(lr){2-3}\cmidrule(l){4-5}
        $r$ & IndexerStage & Memory-Only & Mean & Rank max. \\
        \midrule
        1 & 5.51 & 8.63 & 7.13 & 10.51 \\
        2 & 8.92 & 10.15 & 1.81 & 2.45 \\
        3 & 9.91 & 9.94 & 0.14 & 0.24 \\
        4 & 9.65 & 9.62 & 0.10 & 0.13 \\
        \bottomrule
    \end{tabular*}
    \caption{TPS at the batch sizes in Figure~\ref{fig:eval-admission}. Waiting is the time per decode iteration spent waiting for an execution buffer to become ready. The last two columns give the mean and maximum across ranks for Memory-Only.}
    \label{tab:eval-admission}
\end{table}

Ignoring the transfer budget adds before-scan waiting. At $r=2$, Memory-Only requires 7.06~GB per step against 5.98~GB available, exceeding the budget by 18.06\%. Its mean buffer wait is 1.81~ms; at $r=1$, heavier streaming raises that wait to 7.13~ms. With $r=3,4$, the waits fall to 0.14 and 0.10~ms as transfer pressure eases.

Joint admission keeps planned transfers within the calibrated overlap schedule. At $r=2$, it selects batch size 41 rather than 54 and reduces TPOT by 13.62\%, from 85.20 to 73.60~ms. At $r=3,4$, both checks admit the same batch sizes and TPS differs by at most 0.30\%. Thus the transfer check constrains concurrency where supply would overrun the budget and preserves memory-supported concurrency where transfers fit.

\subsection{Separating the Latent-KV Backing-Path Benefit}
\label{sec:eval-bandwidth}
\label{sec:eval-backing-path}
\label{sec:eval-cost}
\label{sec:eval-overhead}

The larger batch retains a TPS benefit when the latent-KV backing path is held fixed. At 64K, LatentPool-HBM retains Indexer-K in local HBM while sharing IndexerStage's latent-KV access and writeback path. Table~\ref{tab:eval-matched} separates the path-only change from Indexer-K streaming. The operator comparison then explains why additional requests increase TPS less than batch size.

\subsubsection{Execution at Fixed and Larger Batches}
\label{sec:eval-matched-batch}

The latent-KV path change alone gives $1.02\times$ decode TPS at batch size 24 and reduces TPOT by 1.92\%. Both HiSparse and LatentPool-HBM keep Indexer-K resident. The additional $1.10\times$ result therefore comes from enabling a larger batch on an already matched path.

\begin{table}[t]
    \centering
    \footnotesize
    \setlength{\tabcolsep}{2pt}
    \begin{tabular*}{\columnwidth}{@{}l@{\extracolsep{\fill}}rrrr@{}}
        \toprule
        System & \shortstack{Batch\\size} & TPS & \shortstack{TPOT\\(ms)} & \shortstack{TPS /\\control} \\
        \midrule
        HiSparse & 24 & 6,128.10 & 62.66 & $0.98\times$ \\
        LatentPool-HBM & 24 & 6,247.90 & 61.46 & $1.00\times$ \\
        IndexerStage & 24 & 6,100.00 & 62.95 & $0.98\times$ \\
        IndexerStage & 29 & 6,886.70 & 67.38 & $1.10\times$ \\
        \bottomrule
    \end{tabular*}
    \caption{64K TPS and TPOT relative to LatentPool-HBM at batch size 24. Batch size is per rank. HiSparse changes the latent-KV access path; IndexerStage changes Indexer-K placement and execution.}
    \label{tab:eval-matched}
\end{table}

At batch size 24 on the same latent-KV path, Indexer-K streaming adds 2.42\% TPOT overhead. Request count and model kernels are unchanged. The overhead comes from transfer control, readiness checks, writable-page preservation, and shared-resource interference. Completed-page gathering runs in the backing tier, while its writes still use model-serving HBM and interconnect bandwidth.

Increasing batch size to 29 converts the released memory into 20.83\% more tokens per iteration. TPOT grows by 9.63\% relative to the batch-size-24 control, so decode TPS reaches $1.10\times$ the control. The next comparison separates the fixed-batch integration cost from the extra computation and communication required by the additional requests.

\subsubsection{Operator-Level Execution Costs}
\label{sec:eval-operator-cost}

\begin{figure}[!t]
    \centering
    \includegraphics[width=0.88\columnwidth]{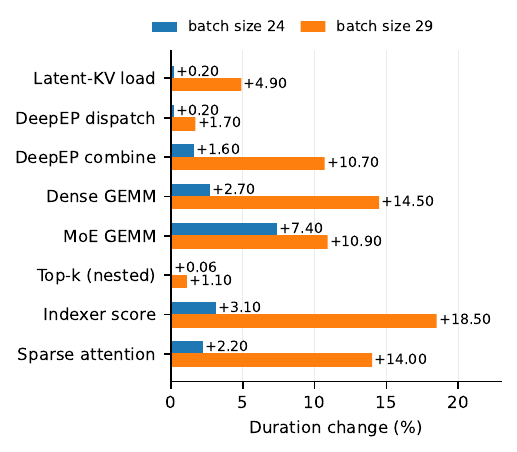}\par
    \caption{Operator-duration changes at 64K relative to LatentPool-HBM at batch size 24. Batch size 24 enables streaming at a fixed workload; batch size 29 adds requests. GEMM denotes general matrix multiplication, and MoE denotes mixture of experts. Indexer score includes top-$k$; overlapping durations are not additive.}
    \label{fig:eval-overhead}
\end{figure}

Figure~\ref{fig:eval-overhead} separates fixed-batch integration overhead from batch-dependent execution. It averages 64 profiled iterations per rank across 16 ranks, using LatentPool-HBM at batch size 24 as the control.

At batch size 24, most operator durations remain close to the control: top-$k$ changes by 0.06\%, and latent-KV loading and expert dispatch by 0.20\%. Indexer score and sparse attention grow by 3.10\% and 2.20\%; expert matrix multiplication has the largest absolute increase, 1.08~ms. These changes reflect streaming integration and resource sharing, not additional requests.

At batch size 29, both state-dependent computation and communication take longer. Indexer score and sparse attention increase by 18.50\% and 14.00\%, while expert combine increases by 10.70\%. More requests require additional scoring and attention work, and their expert outputs must also be combined. This added work explains why TPS grows more slowly than batch size, while the unchanged weights and shared execution costs allow TPS to improve.

\section{Discussion}
\label{sec:discussion}

\paragraph{Choosing how many layers to stream.}
Layer placement determines whether memory or transfers limit the running batch. Section~\ref{sec:eval-admission} shows that streaming more layers can release HBM without increasing admitted concurrency. The additional transfers can require a smaller batch to fit the model execution they are intended to overlap. In the prototype, the layer set is configured before calibration, and admission selects requests under that schedule. A deployment can choose a placement for its desired balance between concurrency and transfer-induced waiting.

\paragraph{Using an existing backing tier.}
IndexerStage can extend a latent-KV pool with storage and transfer capacity for completed Indexer-K pages. The matched-path comparison measures this incremental benefit with the backing hardware held fixed. The prototype uses 16 model-serving and 16 backing H800 GPUs; it is not an equal-cost comparison with an additional model replica. Other backing tiers would need to gather pages outside the model-serving GPU, make completed writes visible before reads, and supply sufficient calibrated bandwidth. A nominal link rate alone does not establish those properties.

\paragraph{Reducing request-update work.}
The dynamic measurements separate background CPU plan construction from the foreground commit that changes execution mappings. Construction overlaps decode under the current plan. Reusing more unchanged mappings could reduce preparation time without changing the native model kernels. Reserving a shorter output-growth horizon could also reduce unused space, provided the runtime renews storage and transfer reservations before admitting that growth. These are extensions to the current bounded-growth policy.

\paragraph{Sharing continuation state.}
Completed latent-KV and Indexer-K pages could form a shared continuation for a future multi-engine deployment. A receiving engine would still need its own joint memory-and-transfer admission and local execution view. Shared storage avoids another persistent copy, but does not remove the bytes needed by each GPU indexer scan. Writable state would require an explicit ownership handoff before the new engine appends entries.

\paragraph{Models that reuse indexer results.}
Cross-layer index reuse reduces the number of layers that run an indexer~\cite{bai2026indexcache}. For such a model, transfer demand includes only offloaded layers whose Indexer-K state is actually scanned. Fewer scans reduce transfer demand, but also change the execution time available for transfers. Applying IndexerStage would therefore require calibrating the new layer schedule rather than assuming that fewer scans automatically make a slower backing path sufficient.

\section{Related Work}
\label{sec:related}

\paragraph{Selective sparse-attention offload.}
ECHO, HiSparse, ESS, and SAC move the latent-KV entries selected for sparse attention~\cite{liu2026echo,hisparse,chen2025ess,ma2026sac}. IndexerStage manages the Indexer-K state used to select those entries. The two kinds of state have different access patterns: attention reads a selected subset, whereas the native indexer scans its full visible context. IndexerStage adds full-state streaming to a latent-KV hierarchy rather than replacing selective attention-data access.

\paragraph{Remote reads and near-memory execution.}
DirectKV eliminates HBM staging buffers by fetching CPU-resident KV tiles inside a redesigned attention kernel~\cite{luo2026no}. Producer and consumer warps overlap fetching and computation within that kernel. KARAT instead runs the indexer and attention on proposed general-purpose near-memory hardware, keeping full Indexer-K scans at the storage node~\cite{ham2026heterogeneous}. IndexerStage keeps the native GPU kernels and stages completed Indexer-K pages before each scan. Its joint admission checks whether the growing batch fits both storage and the calibrated transfer schedule. Memory-Only measures this admission decision within IndexerStage, rather than a fused-fetch or near-memory implementation.

\paragraph{Layerwise and tiered execution.}
SGLang HiCache overlaps loading one layer's cached state with executing another, and vLLM exposes layerwise transfer interfaces~\cite{hicache,vllm-kvconnector,vllm-hybrid-kv}. FlexGen schedules model and attention tensors across GPU, CPU, and storage~\cite{sheng2023flexgen}. These systems provide tiered execution or loading mechanisms. IndexerStage applies layerwise staging to active Indexer-K state that grows during decode. It preserves new entries in local writable storage and reuses execution buffers without changing the native indexer's page-table interface.

\paragraph{Cache pooling and request scheduling.}
Mooncake, LMCache, and TraCT share or transfer KV state across requests and serving engines~\cite{qin2025mooncake,liu2025lmcache,yoon2025tract}. Strata couples cache loading with batch formation and overlaps loading stalls with other work~\cite{xie2026strata}. IndexerStage checks the repeated full-scan demand of active Indexer-K state. Each offloaded page must return before its layer's scan at every decode step. The check includes the whole running batch, output growth, buffer availability, and transfer budgets in layer order.

\paragraph{Reducing indexer computation.}
IndexCache reuses top-$k$ results across layers, while HISA narrows the token search through block filtering followed by token-level refinement~\cite{bai2026indexcache,xu2026hisa}. These methods change how often or over which candidates selection runs. IndexerStage changes state placement while preserving the evaluated native indexer's input and selection semantics. Combining the approaches would require updating page demand and calibrating the resulting execution schedule.

\section{Conclusion}
\label{sec:conclusion}

IndexerStage expands sparse-attention serving capacity beyond latent-KV offload by externalizing completed Indexer-K pages. It preserves the native GPU kernels, reuses local execution buffers across layers, and admits requests against both memory and calibrated transfer budgets. On our H800 testbed, this increases batch size by 12.28--23.08\% and achieves $1.08$--$1.14\times$ the decode TPS of host-memory-backed HiSparse across 16K--64K inputs. Mixed-length dynamic serving achieves $1.31\times$ its decode TPS. Matched-backing-path measurements separate the benefit and execution cost of Indexer-K streaming from changes to latent-KV access.

\label{page:body-end}

\bibliographystyle{plain}
\bibliography{refs}

\end{document}